\documentclass[aps,prl,twocolumn,showpacs,amsmath,amssymb,floatfix,superscriptaddress,
]{revtex4-1}
\usepackage{graphicx}
\usepackage{soul}
\usepackage{mathtools}

\begin{document}
\title{Evidence of a critical phase transition in a purely temporal dynamics with long-delayed feedback}
\date{\today}

\author{Marco Faggian}
\affiliation{SUPA, Physics Department and Institute for Complex Systems and Mathematical Biology, King's College, University of Aberdeen, AB24 3UE (UK)} 
\affiliation{Faculty of Information Studies in Novo Mesto, 8000 Novo Mesto (Slovenia)}

\author{Francesco Ginelli}
\affiliation{SUPA, Physics Department and Institute for Complex Systems and Mathematical Biology, King's College, University of Aberdeen, AB24 3UE (UK)} 

\author{Francesco Marino}
\affiliation{Consiglio Nazionale delle Ricerche, Istituto Nazionale di Ottica, Largo E. Fermi 6, 50125 Firenze (Italy)} 

\author{Giovanni Giacomelli}
\affiliation{Consiglio Nazionale delle Ricerche, Istituto dei Sistemi Complessi, Via Madonna del Piano 10, 50019 Sesto Fiorentino (Italy)} 

\begin{abstract}
Experimental evidence of an absorbing phase transition, so far associated with spatio-temporal dynamics is provided in a purely temporal optical system. A bistable semiconductor laser, with long-delayed opto-electronic feedback and multiplicative noise shows the peculiar features of a critical phenomenon belonging to the directed percolation universality class. The numerical study of a simple, effective model provides accurate estimates of the transition critical exponents, in agreement with both theory and our experiment. This result pushes forward an hard equivalence of non-trivial stochastic, long-delayed systems with spatio-temporal ones and opens a new avenue for studying out-of-equilibrium universality classes in purely temporal dynamics.
\end{abstract}

\maketitle

The concepts of scaling and universality play a prominent role in
statistical physics \cite{Kadanoff90}.  Starting with early -- and
pioneering -- scaling ideas \cite{Widom65, Kadanoff66, HH69} 
up to renormalization group theory \cite{Wilson71, WF72}, they have
been successfully applied to develop a comprehensive theory of critical
phenomena in equilibrium \cite{Yeomans} and, to a large extent in non-equilibrium systems \cite{Tauber17}. Diverging correlation lengths and
a scaling of relevant quantities ruled by universal exponents are the signature of such phenomena. In the framework of spatially extended media, universality is uncovered as the system is inspected at increasing length scales, and often characterized via spatially-resolved measurements of the significant quantities (e.g., correlation functions).

An important question is, to what extent the universality classes predicted and observed in spatio-temporal systems can also hold in a purely temporal dynamics, without explicit spatial degrees of freedom. In this Letter, we bring the first answer to the above issue by investigating the behavior of a stochastic, long-delayed bistable system. 

A delayed feedback sets an infinite-dimensional phase space for the
dynamics \cite{Vogel1965,Hale1977,Farmer1982}. A special case is the so-called long-delay limit, i.e. when the delay time $\tau$ is much longer than any other, internal timescale. Here, a suitable representation \cite{Arecchi1992} permits to unveil the role of the involved multiple time-scales acting as effective spatial variables (for a recent review see \cite{Yanchuk2017}) and a thermodynamic limit is defined as $\tau \to \infty$. This correspondence has been shown to hold in deterministic systems \cite{Giacomelli2012,Larger2013,Larger2015,Javaloyes2015} and even put on rigorous grounds close to a Hopf bifurcation \cite{Giacomelli1996}. However, in the presence of non-trivial stochastic processes only few numerical studies in simple models have been reported \cite{Lopez05, Dahmen2008}. In particular, whether this equivalence is preserved as a critical transition point is approached and correlation lengths diverge has never been tested experimentally.

Here, we address this fundamental question showing that a stochastic, long-delayed bistable laser -- where effective spatial degrees of freedom emerge from different, well separated timescales -- undergoes a genuine out-of-equilibrium active-to-absorbing critical phase transition, belonging to the {\it Directed Percolation} (DP) universality class in one-spatial dimension \cite{HH}.
 
Non-equilibrium models related to epidemic spreading \cite{DP2}, the gravity-driven percolation of fluids through a porous medium \cite{DP1} intermittent, interface-depinning and synchronization phenomena \cite{Jensen, Ginelli2003a, Ginelli2003} and to the transition from laminar to turbulent flows \cite{Pomeau, Goldenfeld,HofPRL} are other well-known examples of this universality class, so far mainly associated to genuine spatiotemporal dynamics. Recently, DP and non-equilibrium phase transitions into absorbing states have been investigated in open many-body quantum systems \cite{Marcuzzi, Everest, Gutierrez}. 

Due to its prominence, DP is commonly regarded as the Ising model of non-equilibrium critical phenomena, but 
experimental evidence has long been elusive. Only recently, measurements in systems displaying spatio-temporal turbulence \cite{Kaz, Hof, Sano2016} have provided the first clear evidence of DP critical behavior
in one and two spatial dimensions, and sparkled renewed attention for this ubiquitous non-equilibrium phenomenon.

\begin{figure}[t!]
\includegraphics[width=1.0\linewidth]{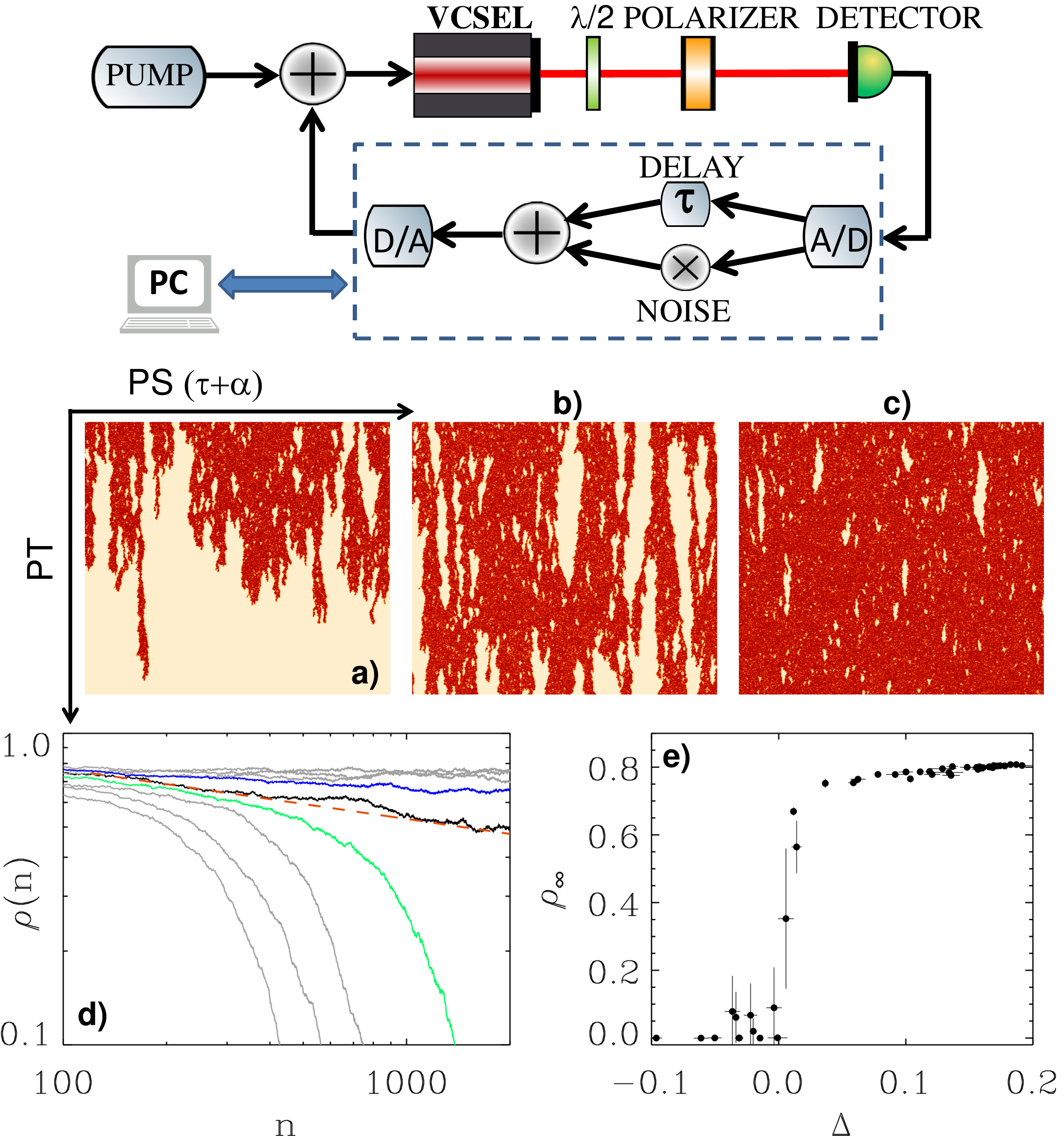}
\caption{Experimental setup (top). Center and bottom: experimental laser dynamics with an upper polarization initial condition, for a fixed feedback gain and multiplicative noise amplitude; the delay is $\tau\!=\!190$ ms. (a)-(b)-(c) P-ST patterns increasing the laser pump current in the absorbed, critical and active phase respectively. Darker colors mark the upper polarized state, the horizontal range is $\tau/10$ and the vertical one spans $2\cdot 10^3$ delay units.  (d) Delay-averaged intensity $\rho(n)$ as a function of PT for different values of the normalized control parameter $\Delta$ (see text). The dashed red line is the power law decay for DP scaling theory. (e) Estimated asymptotic intensity $\rho_\infty$ vs. $\Delta$.}
\label{fig1}
\end{figure}

{\it Experiment.}
The experimental setup (Fig.\ref{fig1}) is based on a bistable Vertical Cavity Surface Emitting Laser (VCSEL) which, for a particular choice of the pump current, displays the coexistence of two linear polarization states of the emitted optical field \cite{Giacomelli1998,VanExter1998}. The two polarizations are separated by means of an half-wave plate and a polarizing beamsplitter and then their intensities are monitored by photodetectors. The signal corresponding to the upper state is then acquired, delayed by a time $\tau=190~ms$ and subsequently fed back into the VCSEL through the pump current using a summer-circuit. A Gaussian noise, software-generated as a sequence of zero-mean, delta-correlated numbers is then multiplied to the (non-delayed) main polarization signal and re-injected as well. Due to its multiplicative nature, this noise vanishes on the lower state, while it affects the other polarization inducing spontaneous fluctuations towards the lower state. The delayed feedback is realized sampling the electric signal from the detector with a A/D-D/A board hosted by a PC driven by a Real-Time Linux OS. The data are treated with a custom software which allow to choose the initial condition, the gain and the amount of multiplicative noise.

We begin our experimental test by a visual investigation of evolution of the polarized intensity signal $I(t)$ (in the following, denoted as intensity).
The transformation $t\!=\!n \tau \!+\! \sigma$ with integer $n$ and real $\sigma \! \in \! \left[0, \tau \right)$
is used to obtain the corresponding Pseudo Spatio-Temporal (P-ST) representation in the plane $(\sigma, n)$. 
The original timeseries is cut in consecutive segments of length $\tau$, each labeled by the {\it pseudo-time} (PT) $n$ and,
inside each slide $\sigma$ marks a {\it pseudo-space} (PS) position in a 1D (one-dimensional) space. 
Due to causality, the information transfer processes are strongly asymmetric in this representation \cite{Yanchuk2017}, yielding a non-zero drift term with velocity (in P-ST units), $\alpha \ll \tau$. For visualization purposes, we adopt the comoving transformation $t\!=\!n' (\tau+\alpha) \!+\! \sigma'$ with $\sigma' \! \in \! \left[0, \tau \!+\!\alpha \right)$ \cite{NOTE0}. 

In this representation, the VCSEL intensity is characterized by nucleation, propagation and annihilation of fronts that, in the absence of multiplicative noise leads to coarsening \cite{Giacomelli2012, Giacomelli2013, Javaloyes2015}. Notably, the P-ST description allows to unfold and display the features of the dynamics over a range of peculiar and independent timescales: the width of the fronts separating the upper and lower polarization states (bandwidth-limited at few $\mu s$), the PS correlation length, the delay $\tau$, and the PT correlation length. All of them play a role in our setup. The ratio between $\tau$ and the fronts width corresponds to the aspect-ratio, as defined in spatially extended systems \cite{Yanchuk2017}. The PS and PT correlation lengths determine the features of the patterns of the active state and are known to scale with the distance from the critical point.

Typical P-ST patterns of the intensity as the pump current is increased and for a fixed multiplicative noise amplitude are shown in Fig.\ref{fig1}a-c. The system is initialized with a sequence of length $\tau$ close to the upper state (with a random Gaussian statistics). We observe a complicated P-ST dynamics with a relaxation towards either the lower (Fig.\ref{fig1}a) or upper (Fig.\ref{fig1}c) state. For intermediate values of the pump, the system apparently evolves slowly on longer PT-scales. 

Such behavior is immediately reminiscent of the active-to-absorbed phase transition, as observed in a large class of reaction diffusion systems 
with an {\it absorbing} state (i.e. a state able to trap the dynamics indefinitely). 
One can readily identify the lower state, preserved by our choice of the multiplicative noise, with such an absorbing state: with other experimental noises carefully minimized, the upper states cannot nucleate spontaneously inside PS homogeneous patches of lower states, which can only be displaced by the invasion of moving fronts. On the contrary, fluctuations can easily nucleate lower state patches inside the upper ones.
Moreover, once the laser sets down in the lower state for at least one full delay, its emission has no more chances to jump back to the upper polarized state, being effectively absorbed. 
 
According to the Janssen-Grassberger conjecture \cite{Janssen1981,Grassberger1982}, in the absence of additional symmetries and/or quenched randomness, any spatio-temporal
dynamics displaying such a transition from a fluctuating {\it active} (i.e. not dynamically frozen) phase into a unique absorbed state is expected to belong to the DP universality class. One should thus expect a critical, power-law behavior described by three independent critical exponents, numerically known with high accuracy in 1D \cite{HH}. 

We use as a control parameter the coarsening velocity $v$ in the absence of the multiplicative noise estimated at the beginning and the end of each measurement. While this velocity is strictly related to the pump current, this procedure allows for a more precise determination of the actual working point of the laser (see Supplemental Material). In particular, we retain only those measurements whose final-initial relative difference in the speed is smaller than $3\%$. As an order parameter, we introduce the delay-averaged intensity $\rho(n) \!=\ \langle I(\sigma,n) \rangle_\sigma$, normalized between $0$ and $1$ and its PT-asymptotic value $\rho_\infty$. 

In Fig.\ref{fig1}d the PT evolution of $\rho(n)$ is reported for different values of the corresponding coarsening velocity. A clear transition is present from asymptotically non-zero values (the active phase) to an exponential decrease towards zero (absorbing phase). The bold green and blue curves correspond to the patterns shown in Fig.\ref{fig1}a and (c) respectively. 
A near-critical curve, displaying a power law decay over more than one decade, is also plotted alongside the known DP asymptotic scaling $\rho(n) \!\sim \! n^{-\delta}$ (with $\delta \!=\! 0.159464(6)$ \cite{HH}), showing a satisfactory agreement between our experiment and DP scaling theory.

We further define the normalized control parameter $\Delta \!=\! (v \!- \! v_c)/v_c$ where $v_c$ is the empirical critical coarsening velocity, and plot the PT-asymptotic value of the order parameter $\rho_\infty$ as a function of $\Delta$. Fig.\ref{fig1}e clearly shows the signature of a transition between the absorbed ($\Delta\!<\!0$) and active phases ($\Delta\!>\!0$). One would now ideally proceed to measure the scaling of $\rho_\infty$ as the critical point is approached from the active phase, $\rho_\infty \!\sim \! \Delta^\beta$ with $\beta \!=\! 0.276486(6)$ \cite{HH}. Unfortunately, when initialized in the upper state the system is prone to non-negligible fluctuations in the working point, which prevented us from reaching the large PTs needed for a clear testing of this latter scaling law.  

\begin{figure}[t!]
\includegraphics[width=1.0\linewidth]{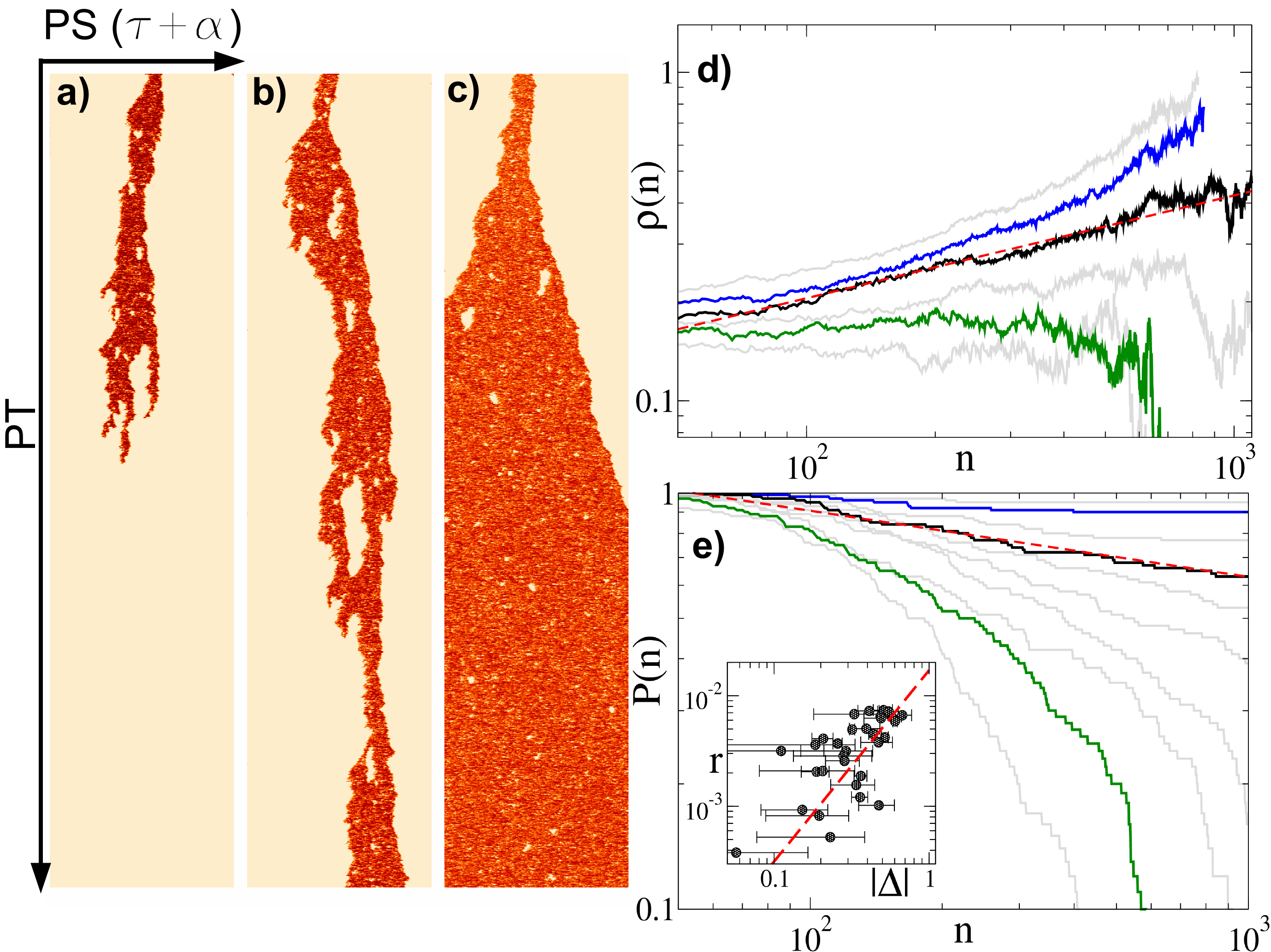}
\caption{Experimental laser dynamics with single seed initial conditions (see text). The delay is $\tau=190$~ms. 
(a)-(b)-(c) P-ST patterns of a single seed in the absorbed, critical and active phase respectively. (d) Delay-averaged intensity $\rho(n)$ and (e) survival probability $P(n)$ as a function of PT for different values of the normalized control parameter $\Delta$ (see text). Inset: survival probability exponential decay rates $r\!=\!1/\xi_\parallel$ vs. $|\Delta|$ in the absorbing phase. The dashed red lines mark the expected DP power law behavior.}
\label{fig2}
\end{figure}

However, we report in Fig.\ref{fig2} the results of another set of measurements: the so called {\it single seed} behavior. For every value of the control parameter, the system is prepared in the initial delay cell close to the lower state, except for a set of small intervals -- equally spaced along the delay length -- which are set in the higher state. This procedure creates an ensemble of $10^2$ active seeds states which evolve independently as long as their ensuing activity remains separated in PS. The relevant quantities are evaluated as averages over such ensembles. Working point fluctuations are milder for (mainly) lower-state initial conditions and we achieve a better control over the critical dynamics.

In DP scaling theory, single seeds initial conditions may decay into the absorbing state or survive and spread with a time-dependent probability 
$P(n)$. In the absorbing phase one expects $P(n) \!\sim \! n^{-\delta} \exp (-n/\xi_\parallel)$, where the temporal correlation length diverges 
as the critical point is approached: $\xi_\parallel \!\sim \! \Delta^{-\nu_\parallel}$ with $\nu_\parallel \!=\! \beta/\delta\!=\!1.733847(6)$
being a second independent exponent  \cite{HH}. Thus at the critical point the survival probability decays as a power law $P(n) \!\sim \! n^{-\delta}$. 
A third independent exponent, the so-called {\it initial slip exponent} $\theta$, can be finally deduced from the initial growth of activity 
at the critical point when starting from a single seed (or sparsely active) initial condition, $\rho(n) \!\sim \! n^{\theta}$ for $\rho(n) \ll 1$, with
$\theta\!=\! 0.313686(8)$ \cite{SLIP}.

In Fig.\ref{fig2}a-c, we present three single-seed sample patterns showing the onset of a near-critical behavior in Fig.\ref{fig2}b. We further report in Fig.\ref{fig2}d-e the growth of the delay-averaged signal (d) and the survival probability (SP) of a seed (e) as a function of PT for different values of $\Delta$. The bold green and blue again denote the sub- and super-critical behaviors corresponding to the patterns in Fig.\ref{fig2}a-c; 
the bold black curve corresponds to the critical case of Fig.\ref{fig2}b. The dashed lines superimposed in the density and SP plots are the power laws 
expected for DP at criticality, showing an excellent agreement with the spatio-temporal theory. In the inset of Fig.\ref{fig2}e, we plot the PT correlation length $\xi_\parallel$ estimated in the sub-critical case. In spite of the large errors due to residual working point fluctuations, the results are compatible with the DP power-law scaling as depicted by the dashed line.

{\it Model.} 
In order to corroborate our experimental findings, we introduce a stochastic effective model, derived from the deterministic description of Ref. \cite{Giacomelli2012},
\begin{equation}
 dx_t = \left[ g\,x_{t-\tau} +F_a(x_t) \right] dt + b \,x_t\, dW_t
 \label{eq:eff}
\end{equation}
where the real variable $x_t$ represents the intensity, $dW_t$ is the increment of a Wiener process, $F_a(x) \!=\!-\frac{d}{dx} U_a(x)\equiv - x(x-1)(x- a)$ a force term derived from a bi-stable quartic potential and $\tau$ the delay
time. The dynamics (\ref{eq:eff}) is controlled by three real parameters, the delayed feedback coupling $g$, the multiplicative noise
amplitude $b$ and the potential asymmetry $a$  with
$a\!>\!g\!>\!0$.

\begin{figure}
\center
\includegraphics[width=1.0\linewidth]{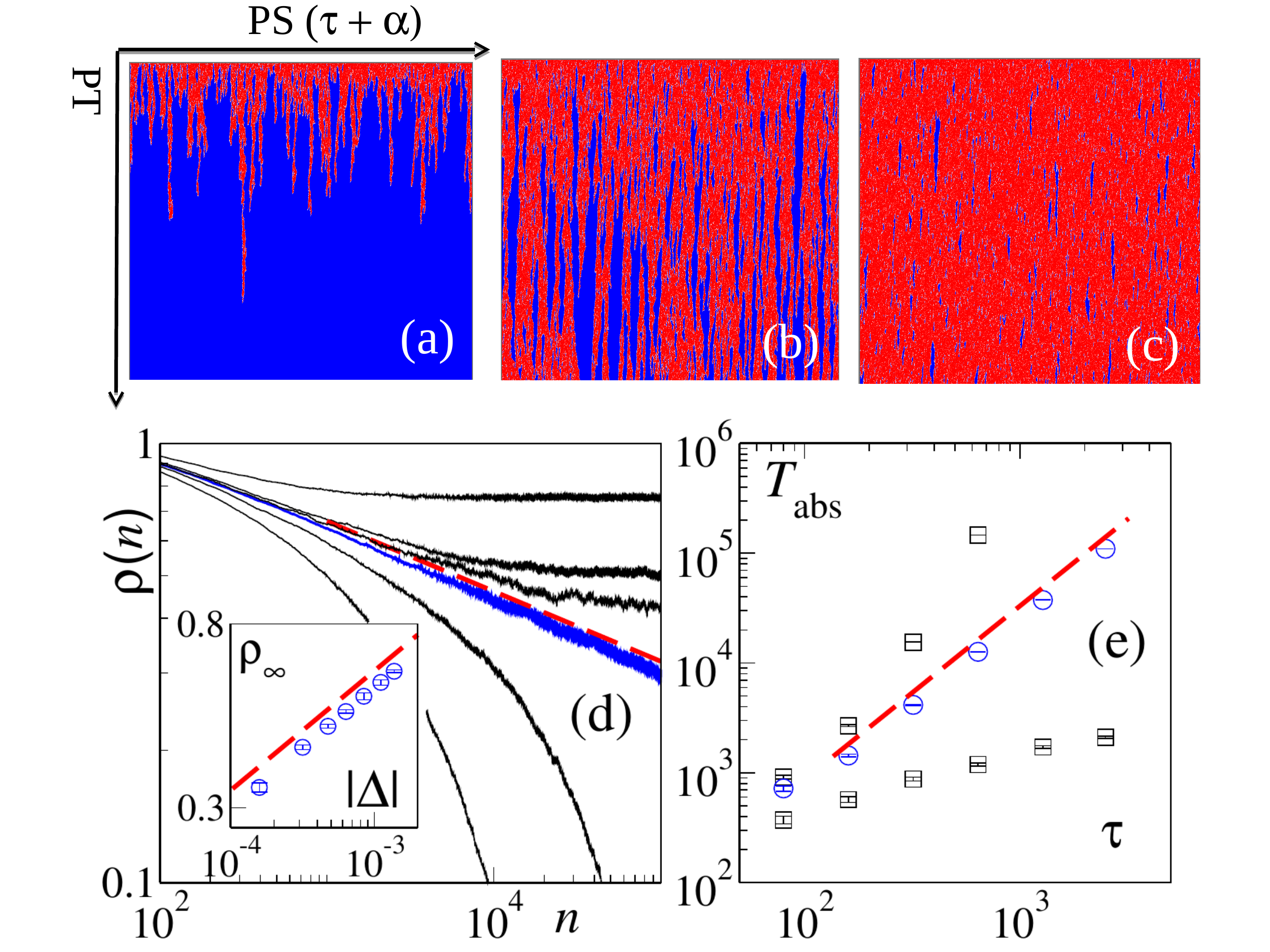}
\caption{Simulations of Eq. (\ref{eq:eff}) for fully active initial conditions. Typical P-ST patterns in the (a) absorbed ($a\!=\!0.93$), (b) critical ($a\!=\!0.89455$) and (c) active ($a\!=\!0.86$) phases in a blue ($x_t \! \approx \! y_0$) to red ($x_t \! \approx \! y_1$) color scale. (d) Delay-averaged intensity $\rho(n)$ for different values of $a$ 
(from top to bottom: $a\!=\!0.88, 0.885, 0.89, 0.89455, 0.8985, 0.9045$).
The delay is $\tau\!=\!2 \cdot 10^5$, the corrected drift $\alpha \!=\! 5.75$. Data has been further averaged over 5 to 50 independent realizations. 
Inset: Scaling of $\rho_\infty$ (estimated by a time-average
of $\rho(n)$ over the stationary regime) vs. $\Delta\!=\!|a-a_c|/a_c$. (e) Mean absorbing time vs. $\tau$  for 
different values of $a$ (from top to bottom: $a\!=\!0.891, 0.8947, 0.887$). Data is averaged over 400 realizations. 
The dashed red lines mark the DP scaling behavior (lower graphs are in log-log scale), while our best fits (not shown) estimate $\delta \! \approx \! 0.16(1)$, $\beta \! \approx \! 0.28(1)$,
$z \! \approx \! 1.58(8)$. }
\label{fig3}
\end{figure}

The deterministic dynamics ($b\!\!=\!\!0$)
has two stable fixed points, $x_t\!=\!y_0\!\equiv\!0$ and
$x_t \!=\! y_1\!\equiv\! [(1\!+\!a) \!+\! \Gamma]/2$, with $\Gamma\equiv \sqrt{(1\!-\!a)^2 \!+\! 4 g}$, separated by the unstable fixed point $x_t \!= y_u\!\equiv\! [(1\!+\!a) \!-\!
\Gamma]/2$. In the absence of delay ($\tau \!\to\! 0$), the deterministic dynamics only consists in a relaxation to equilibria on intrinsic timescales of order $t_0 \!=\! (a\!-\!g)^{-1}$ and $t_1 \!=\! \Gamma^{-1}
(1\!+\!a\!+\!\Gamma)^{-1}$. In the case of a long delay $\tau \!\gg \! t_0,t_1$, quasi-heteroclinic fronts joining the two stable fixed points can be observed as a transient phenomena. Indeed, the relative stability of the fixed points controlled by system parameters, in particular by the potential asymmetry $a$, determines the coarsening dynamics \cite{Giacomelli2012}. 

\begin{figure}
\hspace{-0.5cm}
\includegraphics[scale=.28]{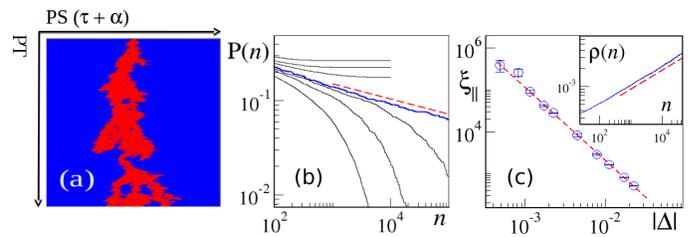}
\caption{Single seed simulations. (a) Characteristic P-ST pattern near the critical point $a_c^{ss} \! \approx \! 0.89447(2)$. (b) Survival probability $P(n)$  for different values of $a$ (from the top: $a\!=\!0.88, 0.885, 0.89, 0.89447, 0.8955, 0.8985, 0.9045$). 
(c) Sub-critical PT correlation length $\xi_\parallel$ (estimated through the fit of $P(n) \!\sim \! n^{-\delta} \exp (-n/\xi_\parallel)$ vs. $\Delta$. 
  Inset: growth of the delay-averaged intensity $\rho(n)$ at the critical point. Data has been averaged over around $10^3-10^4$ different independent realizations.
The dashed red lines mark the DP scaling behaviour, while our best fits (not shown) estimate $\delta \! \approx \! 0.159(4)$, $\nu_\parallel \! \approx \! 1.73(2)$,
$\theta \! \approx \! 0.32(1)$. 
}
\label{fig4}
\end{figure}

In the following, we investigate numerically the full stochastic dynamics for long delays, interpreting equation (\ref{eq:eff}) in the It\^o sense and adopting a simple Euler-Maruyama integration method with a timestepping $\Delta
t\!=\!0.01$ \cite{NOISE}. We fix the noise amplitude $b\!=\!1/\sqrt{7}$ and the delayed
feedback coupling $g\!=\!0.22$, using the potential asymmetry $a$ as our
main control parameter. We have however verified that analogous results
hold using, for instance, $g$ as control parameter. 

We focus on the range $a \in [0.5,1]$, preparing our system with the
initial conditions $x_t \!=\! y_1$ for $t \in [0,\tau )$. Our numerical results for the delay-averaged intensity are reported in the P-ST plots of Fig.~\ref{fig3}a-c.  
For values of the asymmetric parameter $a$ close to 1, the P-ST dynamics
quickly drops from the active state to the absorbing one (we consider a state as absorbed when $x_t<y_u$ for one full delay). As $a$ is lowered, the
system goes through a phase transition located at $a_c \! \approx \! 0.8946(1)$ to reach an active phase. The critical exponents 
$\delta$ and $\beta$ can be estimated within a $6\%$ accuracy as shown in Fig.~\ref{fig3}d.
A third independent exponent can be evaluated by measuring the finite-size scaling of the typical
time $T_{\rm abs}$ needed for a finite size system to be absorbed at the critical point. Scaling theory predicts
$T_{\rm abs} \!\sim \! \tau^z$ with the dynamical exponent $z\!=\!1.580745(1)$, in very good agreement with our numerical simulations
(see  Fig.~\ref{fig3}e).

Single seed simulations, reported in Fig.~\ref{fig4}, further confirm the identification of our stochastic dynamics (\ref{eq:eff}) with the DP universality class. The slight deviation between the size asymptotic
critical point $a_c^{ss} \! \approx \! 0.89447(2)$ and the finite size one reported above for fully active initial conditions is indeed compatible
with the expected PS finite size scaling (see Supplemental Material) \cite{HH}.

{\it Discussion.}
To summarize, we have experimentally shown and numerically confirmed the
existence of a DP critical phase transition in the long-delayed dynamics of a bistable system with multiplicative noise. Our system is purely temporal, and the effective spatial variables involved emerge from the multiple time-scales of the dynamics. 

While the onset of non-equilibrium critical phenomena in systems with long-delay has been previously put forward in simple model systems \cite{Lopez05, Dahmen2008, Lepri1994}, this work represents the first experimental evidence of such behavior. We show that the mapping between long-delayed dynamics and spatio-temporal systems is preserved for stochastic dynamics even as a critical point is approached, so that the former and latter systems may share the same universality class.
  
Our result opens a new avenue for studying experimentally a number of out-of-equilibrium universality classes 
-- for instance the Kardar-Parisi-Zhang class \cite{Pazo2010} -- in purely temporal, long-delayed setups. Moreover, the occurrence of critical phenomena and their scalings in higher effective spatial dimensions could be investigated by means of different types of delayed feedbacks with multiple, hierarchically long delays \cite{Yanchuk2014}. 

Furthermore, absorbing phase transitions such as DP may also take place in non-equilibrium many-body quantum systems \cite{Marcuzzi, Everest, Gutierrez}, for instance, in interacting gases of Rydberg atoms \cite{Marcuzzi2015, Gutierrez}. Moreover, an interesting connection between superradiance in cold atom systems and standard lasing has been recently argued in \cite{Kirton}.  In this context, we expect our system -- where in appropriate conditions quantum fluctuations due to spontaneous emission could have an impact on the transition-- to attract the interest of a larger
community interested in phase transition in quantum simulators.

\begin{acknowledgments}
We wish to thank  S. Lepri and A. Politi for useful discussions.
MF and FG acknowledge support from EU Marie Curie ITN grant n. 64256 (COSMOS).
\end{acknowledgments}


\begin{thebibliography}{24}
\bibitem{NOTE0} The drift velocity $\alpha$ can be evaluated, for instance, by autocorrelation measurements in the timeseries.
\bibitem{SLIP} In one spatial dimension, the initial slip exponent $\theta$ is related to $\beta$ and the 
usual spatial and temporal correlation exponents by the scaling relation $\theta \nu_\parallel \!=\!\nu_\perp \! -\! 2 \beta$ \cite{HH}.
\bibitem{NOISE}
Adopting the Stratonovich interpretation and/or different stochastic integration schemes/timesteppings is not expected
to change the system universality class.
\end{thebibliography}
\end{document}